\documentclass[11pt]{article}
\usepackage{hyperref}
\pdfoutput=1
\begin{document}
\title{3D vortex formation of rigid and flexible plates\\in impulsively starting motion}
\author{Daegyoum Kim and Morteza Gharib \\
\\\vspace{6pt} Graduate Aeronautical Laboratories,\\ California Institute of Technology, Pasadena, CA 91125, USA}
\maketitle
\begin{abstract}
This fluid dynamics video shows three-dimensional vortex formation process for plates in impulsive motion which is investigated experimentally by using defocusing digital particle image velocimetry (DDPIV). Rigid and flexible plate cases are compared in order to study the effect of flexibility on 3D vortex formation and associated hydrodynamic forces. This study was motivated by the general question of how the flexibility of flapping propulsors in flying and swimming animals affects vortex formation and propulsive force. For translating and rotating motion, the flexible plate generates a vortex morphology which is drastically different from that of the rigid plate. We identified the deflection of the tip region as the source of this difference. The flexible plate does not produce a large peak for the hydrodynamic force at the impulsive start and stop. This force trend is correlated with smooth vortex formation and shedding processes.
\end{abstract}
\section{Introduction}
Two videos of high and low quality are
\href{http://ecommons.library.cornell.edu/bitstream/1813/11478/3/APS_movie_mpeg-2-Daegyoum%20Kim%20and%20Morteza%20Gharib.mpg}{Video
1} and
\href{http://ecommons.library.cornell.edu/bitstream/1813/11478/2/APS_movie_mpeg-1-Daegyoum%20Kim%20and%20Morteza%20Gharib.mpg}{Video
2}.

In this video, three-dimensional vortex formation processes by moving plates are demonstrated. Two simple motions of the plate, translating and rotating motions, are considered. In the translating motion case, a thin transparent polystyrene plates is immersed vertically in an aquarium and driven by a traverse. The aspect ratio of the plates is 3.75 (height 150 mm and width 40 mm). The thicknesses of the rigid and flexible plates are 1.65 mm and 0.25 mm respectively. The plate moves with constant velocity 50 mm/sec after impulsive acceleration. Reynolds number is 2000 based on plate width and constant velocity. In the rotating motion case, the same material is used and the aspect ratio is 4 (height 160 mm and width 40 mm). The angle of the attack is 90 degree and the angular velocity is 25 deg/sec. The impulsively starting plate stops after 4 sec. Reynolds number is about 2800 based on plate width and tip velocity. All velocities mentioned here are the velocities of the rigid plate cases. The plates shown in the video are all immersed in water. Instead of mapping the whole flow field near the plate, we focused on the flow near the tip region. 

Iso-surfaces of vorticity magnitude are used to present vortex structures. Due to the tip deformation by flexibility, the vortex morphology of the flexible plate cases is totally different from that of the rigid plate cases. In addition, the vortex forms and sheds smoothly at impulsive start and stop. The influence of smooth vortex formation and shedding is reflected in drag force graphes. In the flexible plate cases, the high peaks of drag are not shown even though the plates start and stop impulsively. The flexibility of the plate is a critical factor in vortex and force generation.

\end{document}